\documentclass[iop,apjl]{emulateapj}
\bibpunct[]{(}{)}{;}{a}{}{,}

\shorttitle{Scattering of helioseismic waves by a sunspot}

\begin{document}

\title{Scattering of helioseismic waves by a sunspot:\\
wavefront healing and folding}

\author{Z.-C. Liang\altaffilmark{1,2}, L. Gizon\altaffilmark{2}, and H. Schunker\altaffilmark{2}}
\affil{
\altaffilmark{1}Physics Department, National Tsing Hua University, Hsinchu 30013, Taiwan (R.O.C.)\\
\altaffilmark{2}Max-Planck-Institut f\"ur Sonnensystemforschung, 37191 Katlenburg-Lindau, Germany; gizon@mps.mpg.de
}

\begin{abstract}

We observe and characterize the scattering of acoustic wave packets by a sunspot, in a regime where the wavelength is comparable to the size of the sunspot. Spatial maps of wave traveltimes and amplitudes are measured from the cross-covariance function of the random wave field. The averaging procedure is such that incoming wave packets are plane wave packets. Observations show that the magnitude of the traveltime perturbation caused by the sunspot diminishes as waves propagate away from the sunspot \,--\, a finite-wavelength phenomenon known as wavefront healing. Observations also show a reduction of the amplitude of the waves after their passage through the sunspot. A significant fraction of this amplitude reduction is  due to the defocusing of wave energy by the fast wave-speed perturbation introduced by the sunspot.  This ``geometrical attenuation'' will contribute to the wave amplitude reduction in addition to the physical absorption of waves. In addition, we observe an enhancement of wave amplitude away from the central path: diffracted rays intersect with unperturbed rays (caustics) and wavefronts fold and triplicate. Thus we find that ray tracing is useful to interpret these phenomena, although it cannot explain wavefront healing.

\end{abstract}


\section{Introduction}

The propagation of solar seismic waves is affected by sunspots. The scattering phase shifts and wave absorption coefficients can be measured by several techniques such as Fourier-Hankel analysis and time-distance analysis \citep[see recent review by][ and references therein]{giz10}.

\citet*{cam08} (hereinafter CGD) studied the interaction of $f$-mode plane-wave packets with a sunspot using time-distance helioseismology. As we shall describe briefly below, the data averaging strategy used by CGD greatly reduces the noise and enables a detailed study of the waveforms far from the scattering region. In this study, we extend the observations to modes with different radial orders and wavelengths that are closer to the sunspot radius, in order to investigate finite-wavelength effects \citep[e.g.][]{nol00,hun01}. Not only do we measure traveltime shifts but also wave amplitude perturbations with respect to the quiet Sun.

We use one-minute cadence Doppler velocity images measured by the Michelson Doppler Imager \citep{sch95} onboard the Solar and Heliospheric Observatory (SOHO/MDI). The images are remapped using Postel's azimuthal equidistant projection in order to track the motion of the sunspot in Active Region~9787 \citep[CGD;][]{giz09,mor10}. The map scale is $0.12$ heliographic degrees per pixel. Nine data sets are obtained, one for each day of the period 20--28 January 2002. The selected sunspot is nearly circular and does not evolve over the duration of the observations;
its outer-penumbral radius is $R=20$~Mm.

\begin{deluxetable}{cccccc}
\tablewidth{0pt}
\tablecaption{Characteristics of wave packets\label{tab:char}}
\tablehead{
\colhead{Modes} & \colhead{$n$} & \colhead{$\nu$ (mHz)} & \colhead{$\lambda$ (Mm)} & \colhead{$R/\lambda$} & \colhead{$x_{\rm F}$ (Mm)}
}
\startdata
$f$   & 0 & 2.6 & \phn6.3 & 3.2 &    126.2\\
$p_1$ & 1 & 2.9 & \phn8.3 & 2.4 & \phn95.3\\
$p_2$ & 2 & 3.1 &    10.7 & 1.9 & \phn73.4\\
$p_3$ & 3 & 3.4 &    13.1 & 1.5 & \phn59.4\\
$p_4$ & 4 & 3.6 &    14.8 & 1.4 & \phn52.2
\enddata
\end{deluxetable}

Filters are applied to the Dopplergrams in 3D Fourier space to extract waves with the same radial order (ridge filters). Five different filters are applied, for modes $f$ through $p_4$ (corresponding to radial orders $n=0$--4). Table~\ref{tab:char} lists the characteristics of the wave packets for each radial order, including the central frequency, $\nu$, and the central wavelength, $\lambda$. Broadly speaking, the finite-wavelength effects are expected to be observable in the regime where $R/\lambda$ approaches unity (this condition will be refined in Section~\ref{sec:heal}). These effects should thus be easier to see with the $p_4$ wave packet.

\section{Cross-covariance function and wave packets}

The cross-covariance of the filtered Doppler velocity is computed in the same way as in CGD:
\begin{equation}
C_n(x,y,t)=\int \overline{\phi}_n(t')\phi_n(x,y,t'+t)\,{\rm d}t',
\label{eq:xc}
\end{equation}
where $(x,y)$ is a local Cartesian coordinate system with origin at the center  of the sunspot, $t$ is the time, $\phi_n$ is the filtered Doppler velocity for modes with radial order $n$, and $\overline{\phi}_n$ is the average of $\phi_n$ over the line $L$ at $x=-43.73$~Mm.  The spacial averaging over $L$ is equivalent to filtering out all the wavevectors that are not perpendicular to $L$: only waves that propagate toward and away from $L$ contribute to $\overline{\phi}_n$.

As explained by \citet{giz10}, for $t>0$, the cross-covariance is used to follow the propagation of two plane wave packets initially at $L$ and propagating toward $+x$ and $-x$ respectively. The interaction of the waves with the sunspot will occur in the $(x>0,t>0)$ quadrant.

To reduce noise, the cross-covariances are averaged over nine days and over azimuthal angle about the sunspot center (see CGD). An example plot of cross-covariances has been given by \citet[][, Figure~17]{mor10}.

\section{Wave traveltimes and amplitudes}
\label{sec:meth}

\begin{figure}
\centering
\includegraphics[scale=1.2]{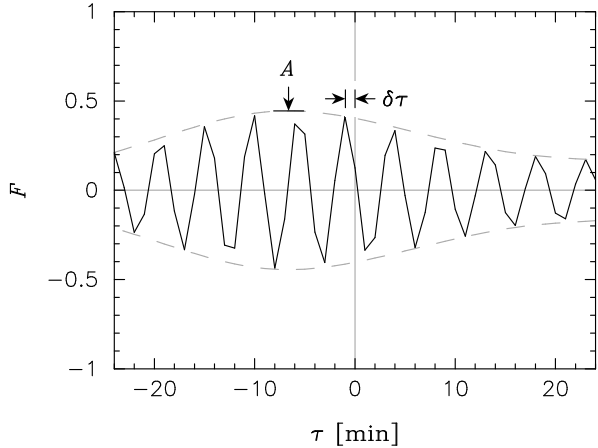}
\caption{
Cross-correlation function $F_4(x,y,\tau)$ (solid line) between the cross-covariance $C_4(x,y,t)$ at $(x,y)=(60,0)$~Mm and the corresponding quiet-Sun cross-covariance $C_4^{\rm qs}$. The envelope (dashed line) is determined by demodulation. The traveltime shift, $\delta \tau$, and the wave packet amplitude, $A$, are indicated by arrows. In practice, we fit the peak nearest to $\tau=0$ with a parabola to measure the time shift. In this particular example, the time shift and reduced amplitude are $\delta \tau=-57$~s and $A=0.44$ with respect to quiet Sun.
\label{fig:fcc}
}
\end{figure}

We define the traveltime shift, $\delta \tau(x,y)$, as the time-lag that maximizes the similarity between the measured cross-covariance, $C_n(x,y,t)$, and a sliding quiet-Sun cross-covariance, $C_n^{\rm qs}(x,t)$. More precisely, $\delta \tau$ is the time $\tau$ that maximizes the function
\begin{equation}
F_n(x,y,\tau)=\frac{\int C_n(x,y,t)C_n^{\rm qs}(x,t-\tau)\,{\rm d}t}{\int |C_n^{\rm qs}(x,t)|^2\,{\rm d}t}.
\label{eq:f}
\end{equation}
The quiet-Sun cross-covariance $C_n^{\rm qs}(x,t)$ is constructed by averaging $C_n(x,y,t)$ over $|y|>100$~Mm, far away from the region of wave scattering, and it is  zeroed out for $t<0$. The cross-correlation traveltime shift defined here is also used in geophysics \citep[e.g.][]{zar10} and is analogous to the definition of \citet{giz02}. The cross-correlation $F_n(x,y,\tau)$ at each spatial point $(x,y)$ as a function of $\tau$ is demodulated using the Hilbert transform  to obtain its envelope.  The maximum, $A(x,y)$, of the envelope of $F_n(x,y,\tau)$ is a measure of the wave packet amplitude. By definition, $A$ is equal to one in the quiet Sun. An example measurement of $\delta \tau$ and $A$ for $C_4$ at $(x,y)=(60,0)$~Mm (after interaction with the sunspot) is shown in Figure~\ref{fig:fcc}.

\begin{figure*}
\includegraphics[scale=1.3]{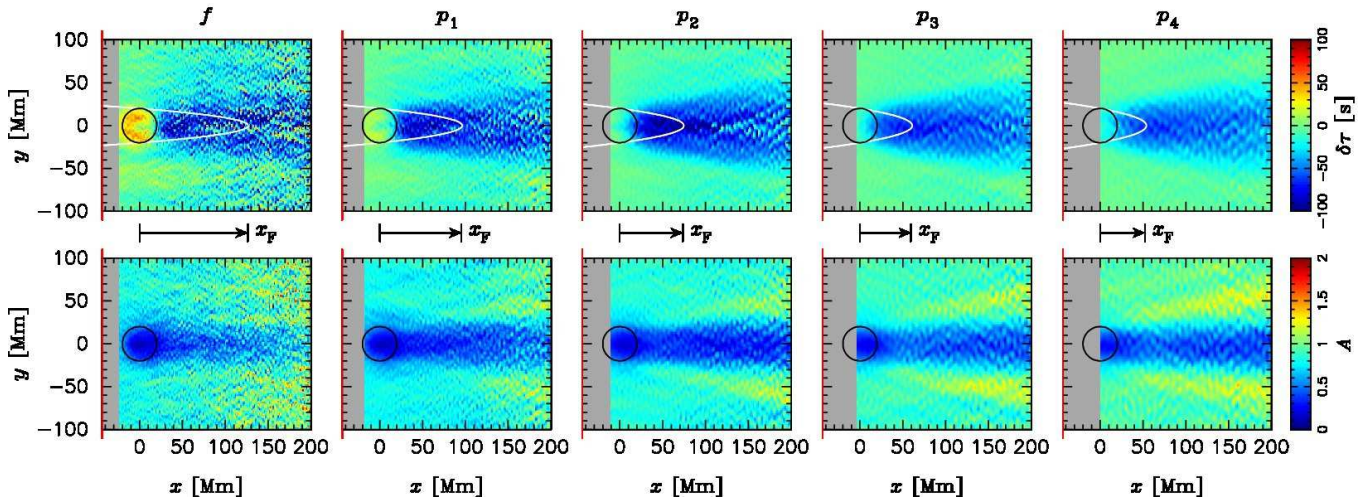}
\caption{
Spatial maps of traveltime shifts $\delta \tau(x,y)$ (top) and wave packet amplitudes $A(x,y)$ (bottom) with respect to quiet Sun. From left to right are results for wavepackets  $f$, $p_1$, $p_2$, $p_3$, and $p_4$. Waves are initially located at $L$ (red lines) and propagate toward $+x$. The black circles denote the outer boundary of the penumbra. The parabolic curves (white lines)  depict the boundaries of the first plane-wave Fresnel zones of a point located at $(x,y)=(x_{\rm F},0)$, chosen such that the sunspot fills the width of the Fresnel zone. We do not display the results in the near fields (gray regions where distances from $L$ are smaller than $3\lambda$).
\label{fig:map}
}
\end{figure*}

For each wavepacket, we obtain spatial maps of the traveltime shift $\delta \tau(x,y)$ and wave amplitude $A(x,y)$, shown in Figure~\ref{fig:map}.  The maps of $\delta \tau$ clearly show negative traveltime anomalies behind the sunspot along $y=0$ while the maps of $A$ show the amplitude reductions.

We only display the results for distances from $L$ larger than three wavelengths in each panel in order to exclude the near field where the $t<0$ and $t>0$ branches of $C_n$ are not well separated. This is not a limitation of this work as we are interested in wave scattering in the far field. For $f$ and $p_1$ modes there are positive traveltime anomalies on the left side of the penumbra, which are likely to be caused by the outward-directed flows in the penumbra and the moat.

\section{Wavefront healing and finite-wavelength effects}
\label{sec:heal}

\begin{figure*}
\centering

\begin{tabular}{cc}
\begin{minipage}[b]{85mm}
\includegraphics[scale=1.25]{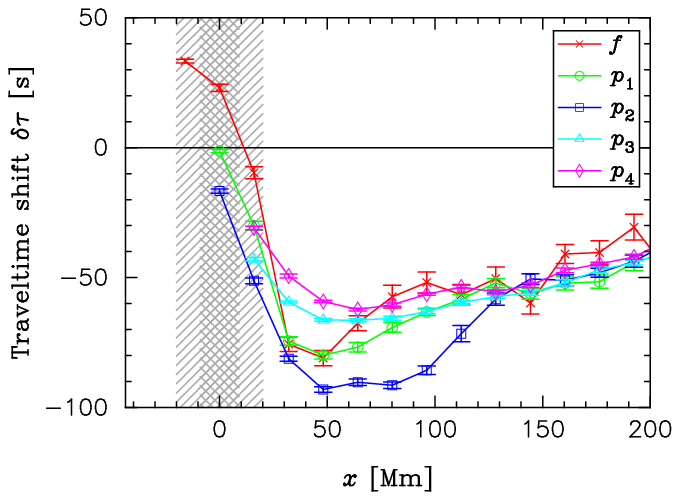}
\end{minipage}
&
\begin{minipage}[b]{85mm}
\includegraphics[scale=1.25]{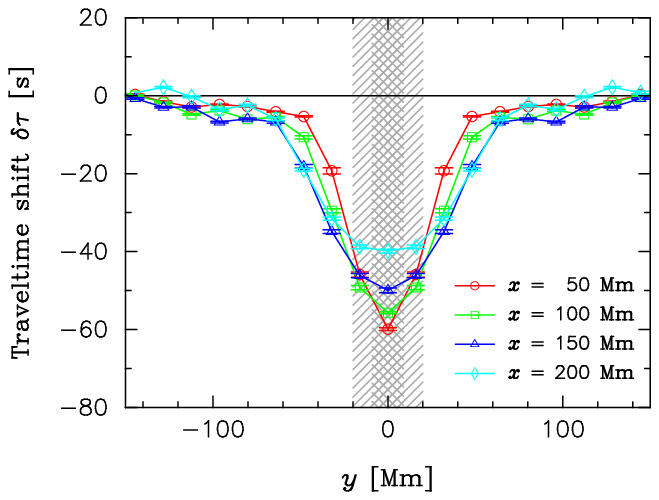}
\end{minipage}
\end{tabular}

\caption{
Observed traveltime shifts, $\delta\tau$. The left panel shows the traveltime shifts along $y=0$ for modes $f$ through $p_4$, which heal as the distance from the sunspot, $x$,  increases. The right panel shows the traveltime shifts of $p_4$ modes as functions of $y$ at four different values of $x$. Notice the spread in the transverse direction as $x$ increases. The hatched and cross-hatched area indicate the locations of the penumbra and umbra respectively. Data are averaged over patches of size $16\times16$~Mm$^2$ and error bars give the standard deviation of the mean.
\label{fig:dt}
}
\end{figure*}

The left panel of Figure~\ref{fig:dt} shows the traveltime shifts as a function of $x$ along $y=0$. For all the modes, the magnitude of the traveltime shift gradually decreases as waves propagate away from the sunspot. This phenomenon, called wavefront healing \citep{nol00,hun01}, is due to the decreasing amplitude of the scattered wavefield with distance. This is a finite-wavelength effect that cannot be explained by ray theory, which would give a constant traveltime perturbation for $x>R$ \citep{giz06}.

The occurrence of finite-wavelength effects at a measurement point depends on whether the diameter of the sunspot is smaller than the width of the first Fresnel zone of that point. For an observation point at $(x,y=0)$ (the point that enters the computation of the cross-correlation), the width (measured along the $y$-axis) of the first plane-wave Fresnel zone is given by  $\Delta y_{\rm F}(x)=2\sqrt{(x+\lambda/4)^2-x^2}$, assuming a path length perturbation of $\lambda/4$ \citep{gud96}. Let us define $x_{\rm F}$, the characteristic $x$ such that the sunspot fills the first Fresnel-zone width, i.e. $\Delta y_{\rm F}(x_{\rm F})=2R$. For each wave packet, the value of $x_{\rm F}$ is provided in Table~\ref{tab:char}. The corresponding first plane-wave Fresnel zone is bounded by a parabola with focus at point $(x_{\rm F},0)$ (white curves in top panels of Figure~\ref{fig:map}). In the regime $x>x_{\rm F}$, the first Fresnel zone is always larger than the sunspot: finite-wavelength effects dominate and the ray approximation is no longer valid.

The right panel of Figure~\ref{fig:dt} shows $p_4$-mode traveltime shifts as a function of $y$, at four different values of $x$. As the distance away from the sunspot increases, we see wavefront healing around $y=0$. In addition, we see that the width of the travel-time perturbation in the transverse direction grows with $x$; this observation cannot be explained in the context of linearized ray theory, according to which travel-time perturbations are computed along unperturbed paths ($y={\rm const.}$) by using Fermat's principle \citep{hun01}. This is another warning against using linearized ray theory in sunspot seismology.

As we shall see in the next section, a proper ray tracing calculation indicates that a fast wave-speed anomaly (like a sunspot) would defocus the rays and lead to travel-time perturbations that spread in the transverse direction as $x$ increases. However, ray tracing will still overestimate the traveltime shifts in the far field and would be incapable of explaining wavefront healing along $y=0$.

\section{Amplitude enhancements and caustics}

\begin{figure*}
\centering

\begin{tabular}{cc}
\begin{minipage}[b]{90mm}
\includegraphics[scale=1.25]{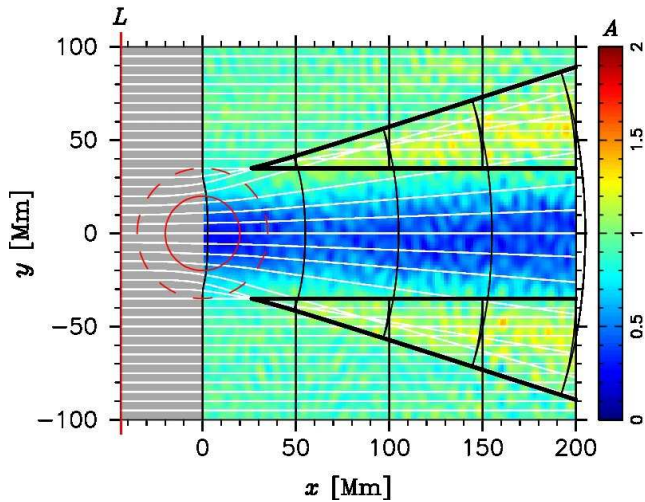}
\end{minipage}

&

\begin{minipage}[b]{85mm}
\includegraphics[scale=1.25]{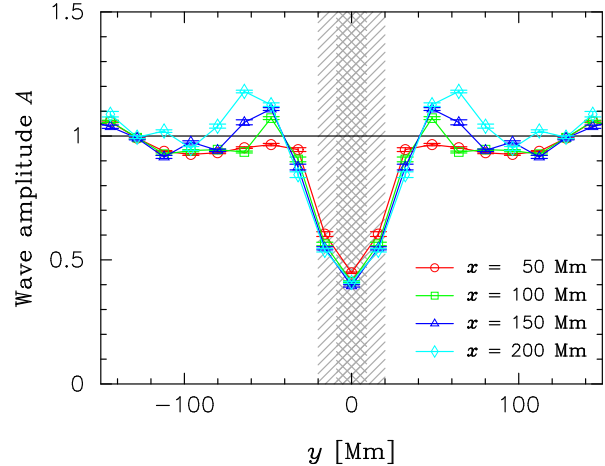}
\end{minipage}

\end{tabular}

\caption{The wave amplitude perturbations for $p_4$ caused by the sunspot.
\emph{Left}: Map of observed amplitudes $A$ (color scale). The white lines are 2D rays traced through a toy sunspot model, such that the wave speed is enhanced by 10\% inside the circle and transitions smoothly to the background value ($\nu\lambda=53.28$ km s$^{-1}$) at the dashed circle. The wavefronts are given by the black lines. The thick black lines indicate the boundaries of the causics where rays intersect and wavefronts fold.
\emph{Right}: Observed amplitudes $A$ as a function of $y$ at fixed distances from the sunspot ranging from $x=50$ to 200~Mm. Notice the amplitude enhancements around $y\sim \pm60$~Mm.
\label{fig:A}
}
\end{figure*}

The bottom panels of Figure~\ref{fig:map} show that the transmitted wave packets have a reduced amplitude around $y=0$ compared to the quiet-Sun value, for all radial orders. It is known that wave absorption by sunspots causes a reduction in outgoing wave amplitude \citep{bra87,bra88}, as the result of partial mode conversion of incoming waves into slow magnetoacoustic waves that propagate down the sunspot \citep[e.g.][]{spr92,cal97}.

In our case (plane wave geometry), a significant fraction of the amplitude reduction that is observed on-axis behind the sunspot is  due to the defocusing of wave energy by the fast wave-speed perturbation introduced by the sunspot.  This can be illustrated by a simple 2D ray tracing experiment shown in the left panel of Figure~\ref{fig:A}. This ``geometrical attenuation'' will contribute to the wave amplitude reduction in addition to the physical absorption of waves. Their respective contributions would require better modeling as ray tracing cannot take account of wavefront healing.

Apart from the on-axis amplitude reduction, an enhancement of wave amplitude is seen away from the axis $y=0$ for modes $p_2$, $p_3$, and $p_4$ (bottom panels in Figure~\ref{fig:map}). This off-path enhancement is more pronounced for the $p_4$ wave packet (right panel of Figure~\ref{fig:A}). Like above, we suggest that this off-path feature is caused by sunspot-induced refraction.  Since rays tend to bend toward the slow-speed medium, diffracted rays will cross over the unperturbed rays on both sides off the central axis. The envelope of the bundle of intersected rays is called a caustic in optics. The wavefronts fold and triplicate as they pass through the caustics. Inside the caustics, all three arrivals will contribute to cross-correlation traveltime measurements and the interpretation of traveltime becomes ambiguous under the ray-theoretical picture \citep{nol00,hun01}. The location of the caustics depends on the perturbation strength and the sunspot geometry. In the case of $p_4$ modes, a 10\% wave-speed perturbation provides a rather satisfying explanation for the on-axis power deficit and off-path enhancement (left panel of Figure~\ref{fig:A}). We note that the off-path enhancements are also seen in the cross-covariance power maps (not shown here), though not as clearly.

\section{Discussion}

We have measured the traveltime shifts and amplitudes of plane wave packets as they traverse a sunspot using a  cross-covariance technique. Observations show on-axis wavefront healing and amplitude reduction. Traveltime anomalies spread in the transverse direction with distance and amplitude enhancements are seen away from the central ray.

Wavefront healing is a finite-wavelength phenomenon. The only other direct observational evidence of finite-wavelength effects in time-distance helioseismology was reported by \citet{duv06} who studied the interaction of waves with sub-wavelength magnetic features. Wavefront healing cannot be modelled by ray theory. This implies that ray-based interpretations are not always appropriate in sunspot seismology.

We have seen that ray-tracing (unlike linearized ray theory) is quite useful to study the geometrical attenuation due to the defocusing of wave energy by a fast wave-speed anomaly. We conclude that wave amplitude reduction is not necessarily due to wave absorption by the sunspot.  Ray tracing is also useful to interpret off-path amplitude enhancements due to wavefront folding, although it cannot explain wavefront healing.

The observations of traveltime shifts and wave amplitude that have been presented in this paper contain a wealth of information about the  seismic signature of sunspots. Their interpretation requires methods of analysis more sophisticated than linearized ray theory (which is still often used in sunspot seismology today).

\acknowledgements
Z.C.L. would like to thank the helioseismology group at the Max-Planck-Institut f\"ur Sonnensystemforschung for their hospitality and D.-Y. Chou for his advice. Z.C.L. was supported by the NSC of ROC under the grants NSC98-2917-I-007-121 and NSC96-2112-M-007-034-MY3.
This study is supported by the European Research Council under the European Community's Seventh Framework Programme (FP7/2007-2013)/ERC grant agreement \#210949,  ``Seismic Imaging of the Solar Interior'',  to PI L. Gizon (Milestone \#4).
SOHO is a project of international cooperation between ESA and NASA.


\end{document}